# Increasing Flips per Second and Speed of p-Computers by Using Dilute Magnetic Semiconductors to Implement Binary Stochastic Neurons

Rahnuma Rahman and Supriyo Bandyopadhyay

*Department of Electrical and Computer Engineering, Virginia Commonwealth University, Richmond, VA 23284, USA*

*Abstract*— Probabilistic computing with binary stochastic neurons (BSN) implemented with low- or zero-energy barrier nanoscale ferromagnets (LBMs) possessing in-plane magnetic anisotropy has emerged as an efficient paradigm for solving computationally hard problems. The fluctuating magnetization of an LBM at room temperature encodes a p-bit which is the building block of a BSN. Its only drawback is that the dynamics of common (transition metal) ferromagnets are relatively slow and hence the number of uncorrelated p-bits that can be generated per second – the so-called "flips per second" (*fps*) – is insufficient, leading to slow computational speed in autonomous co-processing with p-computers. Here, we show that a simple way to increase *fps* is to replace commonly used ferromagnets (e.g. Co, Fe, Ni), which have large saturation magnetization $M_s$, with a dilute magnetic semiconductor like GaMnAs with much smaller saturation magnetization. The smaller $M_s$ reduces the energy barrier within the LBM and increases the *fps* significantly. It also offers other benefits such as increased packing density for increased parallelization and reduced device to device variation. This provides a way to realize the hardware acceleration and energy efficiency promise of p-computers.

*Index Terms*—Probabilistic computing, p-bits, low-barrier-nanomagnets, flips per second, dilute magnetic semiconductors

## I. INTRODUCTION

PROBABILISTIC computing with binary stochastic neurons (BSN) has emerged as a mainstream approach to solving computationally demanding problems such as integer factorization [1] and implementing challenging hardware such as reversible AND logic [2]. A BSN can be physically implemented with a low (or zero) energy barrier nanoscale ferromagnet (LBM) possessing in-plane magnetic anisotropy [3]. The magnetization vector of an LBM, which is usually made of a circular or nearly circular magnetic disk, fluctuates randomly in time at room temperature owing to thermal perturbations and the probability distribution of the fluctuating magnetization can be controlled by injecting a spin polarized current into the LBM or by subjecting it to (voltage generated) uniaxial stress. This allows for a very convenient rendition of a p-bit with controlled distribution.

A p-computer made of LBMs connected with synapses can be operated either with a clock (sequential processing) or without (autonomous co-processing) [4, 5]. In the clocked operation, the computational speed $f$ is limited to $f \leq N/\tau_{clock}$ where $\tau_{clock}$ is the clock period and $N$ is the number of BSNs that can be updated simultaneously (during execution of an algorithm) without sacrificing fidelity of the solution [5]. In this case, the clock speed $1/\tau_{clock}$ sets the limit on the computational speed and the BSN speed does not play any limiting role.

On the other hand, in *clockless* or autonomous co-processing, the computational speed is not constrained by a clock speed but only by the speed of the binary stochastic neurons (BSN) implemented with the LBMs and the speed of synapses [5]. Autonomous co-processing allows for much faster processing than sequential processing with a clock. The BSN/LBM speed is the *number of uncorrelated p-bits that can be generated per second*, or the so-called "flips per second" (*fps*) [5-7]. The computational speed is limited to $f \leq N_0 \times fps$ [where $N_0$ is the number of BSNs], which immediately tells us that a larger *fps* is beneficial. Generally, the resources (time, energy) needed to solve a problem in autonomous co-processing scales inversely with the *fps*. It has been claimed that the *fps* is a problem and substrate independent figure of merit for hardware annealers such as Ising machines where a given problem is solved with autonomous probabilistic co-processing [5].

It is therefore of utmost importance to devise strategies to increase the *fps* associated with LBMs that are used as BSNs. Ferromagnetic LBMs with in-plane anisotropy have much faster dynamics than those with perpendicular anisotropy [3, 8], but their *fps* is still only ~$10^7$ [8, 9] if they are made of transition metals like Co, Ni or Fe. This small *fps* makes the time to solve a problem unduly long since typically many p-bit samples are needed to solve a problem correctly [2]. Antiferromagnets are likely to have much larger *fps*, but their magnetizations, if any, are small (at least for collinear antiferromagnets) which makes sensing the p-bit very difficult.

In this Letter, we offer a solution to this conundrum, whereby we can still use a ferromagnet with a relatively large magnetization (much larger than that of antiferromagnets, even non-collinear ones) which makes sensing not difficult at all and yet increase the *fps* by more than an order of magnitude. In a ferromagnetic nanomagnet, the *fps* is proportional to the hopping rate $1/\tau$ which is the rate at which the magnetic state tries to hop over the internal energy barrier Δ within the nanomagnet. This rate is given by the Arrhenius relation $1/\tau = (1/\tau_0)e^{-\Delta/kT}$ where $\tau_0$ is the attempt time and Δ is the energy

Corresponding author: Supriyo Bandyopadhyay (sbandy@vcu.edu).

barrier [10, 11]. The above relation is, of course, not exact when $\Delta < kT$ but nevertheless, roughly speaking, the *fps* should increase with decreasing energy barrier height. A ferromagnetic LBM with a smaller energy barrier is expected to produce a larger *fps*.

In a slightly elliptical nanomagnet with in-plane shape anisotropy and no magneto-crystalline anisotropy (since the nanomagnets are amorphous), the in-plane energy barrier is given (within the macrospin approximation) by

$$\Delta = \frac{\mu_0}{2} M_s^2 \Omega \left( N_{min} - N_{maj} \right), \quad (1)$$

where $\mu_0$ is the permeability of free space, $\Omega$ is the nanomagnet volume, $N_{min}$ and $N_{maj}$ are the demagnetization factors along the minor and major axes of the ellipse, respectively, and $M_s$ is the saturation magnetization. The expressions for the demagnetization factors in the case of a slightly elliptical cross-section are [12]

$$N_{min} = \frac{\pi}{4}\left(\frac{t}{a}\right)\left[1 + \frac{5}{4}\left(\frac{a-b}{a}\right) + \frac{21}{16}\left(\frac{a-b}{a}\right)^2\right],$$

$$N_{maj} = \frac{\pi}{4}\left(\frac{t}{a}\right)\left[1 - \frac{1}{4}\left(\frac{a-b}{a}\right) - \frac{3}{16}\left(\frac{a-b}{a}\right)^2\right] \quad (2)$$

where $a$ = major axis dimension, $b$ = minor axis dimension and $t$ = thickness of the elliptical nanomagnet. For a perfect circle, $a = b$ and hence $\Delta = 0$. For an ellipse, $a > b$ and hence $\Delta > 0$.

It is obvious from Equation (1) that for a given cross-sectional shape, one way to reduce $\Delta$ is to *reduce the saturation magnetization* $M_s$ via proper choice of materials. A material like Co has a saturation magnetization of $8 \times 10^5$ A/m, whereas a dilute magnetic semiconductor like GaMnAs has a saturation magnetization of only $5 \times 10^3$ A/m [13] at room temperature, which is more than two orders of magnitude smaller. GaMnAs is ferromagnetic at room temperature [13]. Hence, everything else being the same, a GaMnAs LBM will have an energy barrier that is roughly four orders of magnitude smaller than that of a Co LBM. Therefore, it should ideally exhibit an *fps* that is much larger.

GaMnAs offers three other important properties that are extremely beneficial: (1) a magnetic tunnel junction (MTJ) is likely to be used to "read" the magnetization of the LBM and hence a large tunneling magneto-resistance (TMR) would be helpful. Fortunately, GaMnAs/AlAs/GaMnAs MTJs with TMR of up to 75% have been reported [14] and a large TMR has been predicted for GaMnAs/GaAlAs/GaMnAs MTJs as well [15]. Hence, the readout ability is not compromised if Co or CoFeB is replaced by GaMnAs. (2).

Second, the dipole interaction energy at any instant of time $t$ between two LBMs denoted as $i$ and $j$ is given by [12]

$$E_{ij}(t) = \frac{\mu_0 M_s^2 \Omega^2}{4\pi |\vec{r}_{ij}|^3}\left[\vec{m}_i(t) \bullet \vec{m}_j(t) - \frac{3}{|\vec{r}_{ij}|^2}\left(\vec{m}_i(t) \bullet \vec{r}_{ij}\right)\left(\vec{m}_i(t) \bullet \vec{r}_{ij}\right)\right] \quad (3)$$

where the $\vec{m}(t)$-s are the magnetizations of the two LBMs normalized to the saturation magnetization at any instance of time $t$ and $\vec{r}_{ij}$ is the separation between the two LBMs. The strength of the dipole interaction is also proportional to $M_s^2$, which means that it is also more than four orders of magnitude smaller in GaMnAs than in, say, Co. As a result, GaMnAs LBMs can be much more closely packed together than Co LBMs without generating unwanted correlations between p-bits. This will enable massively parallel p-bit architectures which will vastly decrease the time to a solution [16].

Third, GaMnAs provides less device to device variation in LBMs (due to defects and imperfections) compared to a transition metal ferromagnet like Co [17]. This is also related to the lower saturation magnetization [17]. Device to device variation can result in significant spread in the *fps*, limiting the actual operating speed to the smallest of the *fps*-s. Equally important, the statistics (e.g. the full-width-at-half-maximum of the auto-correlation function of the magnetization fluctuation) is much more independent of the *initial condition* (initial orientation of the magnetization) in a GaMnAs LBM than in a Co LBM. In the latter, the auto-correlation function has a strong dependence on the initial magnetization orientation [17].

Lastly, GaMnAs is a well-known and well-researched dilute magnetic semiconductor. It is not significantly different from GaAs and there is a large body of work on integrating GaAs with Si technology [18]. Hence, integrating GaMnAs based p-bit generators or BSNs with CMOS in a Si platform should not pose insurmountable challenges.

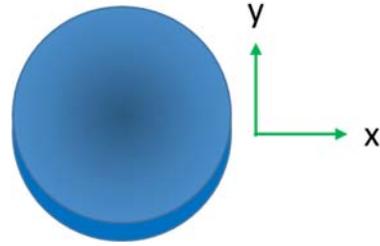

Fig. 1: A slightly elliptical disk of a magnetic material with in-plane anisotropy used as a low barrier nanomagnet (LBM). It acts as a building block for p-bits.

## II. RESULTS

To study the *fps* of transition metal ferromagnets and dilute magnetic semiconductors, we simulate the room-temperature magneto-dynamics of two sets of slightly elliptical nanomagnets, one made of Co and the other made of GaMnAs, using the Landau-Lifshitz-Gilbert equations and three different initial conditions (i.e. different initial magnetization orientations). This yields the magnetization vector as a function of time and allows us to assess the associated *fps*. We consider nanomagnets that are slightly elliptical with major axis of 100 nm and minor axis of 99.9 nm. The magnet thickness is 15 nm. The Gilbert damping factor in GaMnAs is 0.01 [19].

Fig. 1 shows the cross section of a nanomagnet whose major axis is along the *y*-axis and minor axis is along the *x*-axis. The three initial magnetizations are: (1) very close to the major axis $\left[m_y(0) = -0.998, m_x(0) = 0.001\right]$, (2) very close to the minor axis $\left[m_x(0) = -0.998, m_y(0) = 0.001\right]$, and (3) $45^0$ to either axis $\left[m_x(0) = m_y(0) = 0.707\right]$.

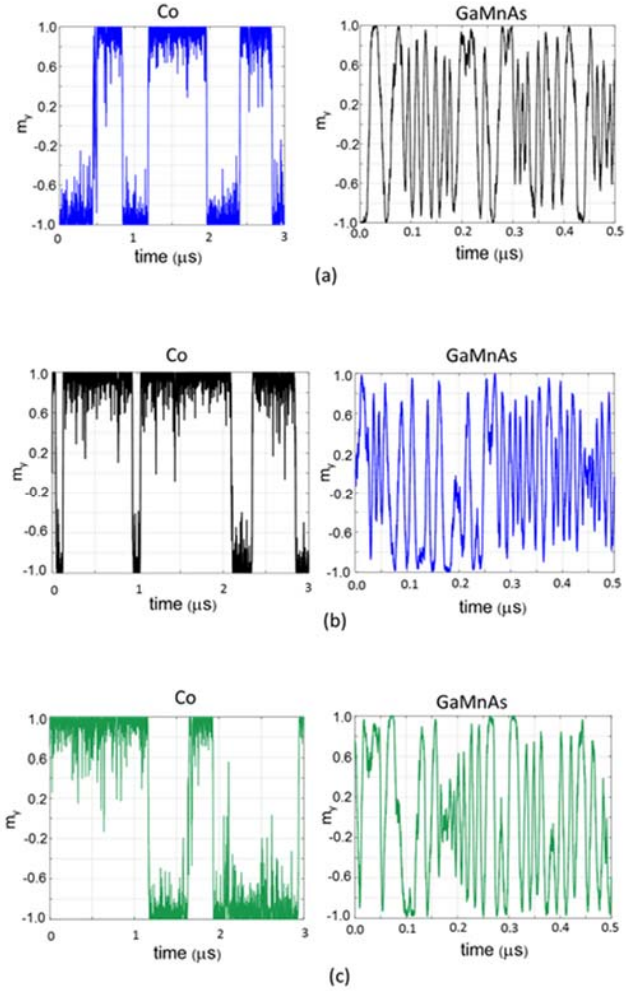

Fig. 2: Fluctuations in the magnetization component along the major axis ($m_y$) for Co and GaMnAs LBMs. Initial magnetization (a) along the major axis, (b) along the minor axis and (c) at $45^0$ to both axes.

In Fig. 2, we show the time dependent fluctuations in the component of the magnetization along the major axis of the elliptical LBM for the three different initial conditions and for both Co and GaMnAs LBMs. For Co, $fps \sim 10^6 - 5\times 10^6$, whereas for GaMnAs, $fps \sim 5 \times 10^7 - 10^8$. Thus, LBMs made of dilute magnetic semiconductors, which have much smaller saturation magnetization than common transition metal ferromagnets, can significantly increase the flips per second and speed up the execution of probabilistic algorithms with p-bits.

### III. CONCLUSIONS

Here we have shown a simple strategy to increase the flips per second (*fps*) in an LBM with an appropriate material choice, i.e. by replacing common transition metal ferromagnets with dilute magnetic semiconductors.

It is interesting to note that the *fps* does not quite scale exponentially with the negative of the barrier height, or the negative of the square of the saturation magnetization, in accordance with the Arrhenius relation. The scaling is much weaker than that, but this is expected since the Arrhenius relation does not hold when the activation energy (in this case the shape anisotropy barrier height) is less than the thermal energy kT. The energy barrier in the case of GaMnAs is less than $10^{-4}$ kT in our example, and hence the exact Arrhenius relation is not expected to hold, but the trend holds up, i.e. the *fps* does increase with decreasing energy barrier height due to decreasing saturation magnetization.

Replacing transition metal ferromagnets with dilute magnetic semiconductors with much smaller saturation magnetization offers other benefits as well, such as less dipole interaction among LBMs that allows increasing packing density to enable massively parallel architectures for high speed, as well as reduced device to device variations and reduced sensitivity to initial conditions.

### ACKNOWLEDGMENT

This work is supported in part by the National Science Foundation under grant CCF-2006843.